\input epsf
\documentstyle[12pt]{article}
\begin{document}
\begin{titlepage}
\setcounter{page}{0}
\rightline{Preprint YERPHI-1448(18)-95}
\vspace{2cm}
\begin{center}
{\Large HIGGS BOSON MASSES IN SUPERSYMMETRIC THEORIES}  \\
\vspace{1cm}
{\large ASATRIAN H.M., YEGHIYAN G. K.} \\

\vspace{1cm}
{\em Yerevan Physics Institute, Alikhanyan Br. 2, Yerevan, Armenia}\\
{\em e-mail: "hrach@uniphi.yerphi.am"}\\
\end{center}
\vspace{5mm}

\centerline{{\bf{Abstract}}}

The Higgs boson mass problem is considered in the next to minimal
supersymmetric standard model. The Higgs potential and the
renormalization group equations for
the gauge, Yukawa and scalar coupling constants are analyzed.
The restrictions for the Higgs boson masses are found
for the cases of presence and absence of spontaneous CP- violation.
\vfill
\centerline{\large Yerevan Physics Institute}
\centerline{\large Yerevan 1995}

\end{titlepage}
\newpage

1.The aim of our paper is to consider the problem of Higgs masses
in the next to minimal supersymmetric standard model (NMSSM).
Such a model contains an additional Higgs singlet, as compared
with minimal supersymmetric standard model (MSSM).
The cause of the consideration of the NMSSM  was connected,
in particular, with desire to avoid the
explicit mass term for the Higgs doublets in the superpotential
\cite{1}.   On the other
hand it is interesting to verify, if the predictions of MSSM
remain valid for the more general supersymmetric models.
The other interesting problem is connected with CP -violation:
it is known \cite{2} that in MSSM one can't realize the realistic
scenario of spontaneous CP-violation, therefore it is natural
to consider, if this scenario possible in NMSSM \cite{3}.
The spontaneous CP- violation leads, in particular,
to the existence of large (compared with the standard model predictions)
electric dipole moment of fermions \cite{3,4,5}.

2.We will investigate the  restrictions for Higgs boson masses
which can be obtained from the Higgs potential and renormalization group
equations in NMSSM.
The  restrictions for Higgs boson masses in NMSSM was investigated
in \cite{6,7,8,9} for the case of absence of spontaneous CP violation.
Here the case of the spontaneous CP- violation will be also considered.
The problem of possible restrictions which can be obtained for
particles masses from the renormalization group equations
analysis in
the case of (nonsupersymmetric) standard model also was considered
in literature.
There various approaches have been developed \cite{10,11,12}.
Such a investigations have been carried out also for the SUSY
extensions of the standard model (see \cite{13} and references therein).

In our consideration
we assume that all of the ( gauge, Yukawa, scalar) coupling constants
of our theory are small between weak ($\sim 100GeV$) and
unification scales ($\sim (10^{16}-10^{18})GeV$), so that the
perturbation theory applies \cite{11,14}. Besides this, for the theory to be
a correct one, the condition of vacuum stability is necessary.
These conditions leads to restrictions for parameters of the
scalar potential. As a result, one obtains some bounds for
Higgs particles masses.

3.Let as proceed to description of our model.
The Higgs sector consists of two Higgs doublets:
\begin{displaymath}
H_{1} =
\left(
\begin{array}{c}
\xi_{1}^{+} \\
\xi_{1}^{o}
\end{array} \right) \\
\hspace{0.5cm}
H_{2} =
\left(
\begin{array}{c}
\xi_{2}^{o} \\
\xi_{2}^{-}
\end{array} \right)
\end{displaymath}
with hypercharges $Y(H_{1})=1, Y(H_{2})=-1$ and the complex
$SU(2)_{L} \times U(1)_Y$ singlet N. The superpotential for
one quark generation is the following
\cite{5}:
\begin{equation}
W = \frac{\lambda_{1}}{3}N^{3} + \lambda_{2}NH_{1}H_{2} +
    h_{u}H_{2}Q_{L}^{c}u_{R} +
h_{d}H_{1}Q_{L}^{c}d_{R}                                       % (1)
\end{equation}
where
\begin{displaymath}
Q_{L} =
\left(
\begin{array}{c}
u_{L} \\
d_{L}
\end{array} \right) \\
\hspace{0.5cm}
Q_{L}^{c} = i\sigma_{2}Q_{L}^{*}
\end{displaymath}
The scalar potential is given by:
\begin{equation}
V = \Sigma|\frac{\partial W}{\partial \varphi_{i}}|^{2} +
\frac{1}{2}\Sigma|g_{a}\varphi_{i}^{+}T_{a}\varphi{i}|^{2} + V_{soft}     % (2)
\end{equation}
where $\varphi_{i}$ (i=1,2...) are the scalar components of all
superfields, $g_{r}$ and $T_{r}$ (r=1,2,3) are coupling constants
and generators of gauge groups
$U(1)_{Y}$, $SU(2)_{L}$, $SU(3)_{c}$  respectively \cite{15,16}.
The term $V_{soft}$ is the part
of potential which breaks the supersymmetry.
It is usually assumed that supersymmetry breaking occurs in
the hidden sector of N=1 supergravity theories \cite{17,18,19,20}.
For the simplest N=1 supergravity models the structure of
supersymmetry breaking terms is:
\begin{equation}
V_{soft} = AmW_{3} + BmW_{2} + m^{2}\Sigma|\varphi_{i}|^{2} +
M\lambda_{r}\lambda_{r}+ h.c.                              % (3)
\end{equation}
where m is the gravitino mass, A, B are supersymmetry breaking
parameters, $\lambda_{r}$ (r=1,2,3) are gaugino fields and M is gauginos
universal mass.

In the present paper we will consider more general supergravity
models \cite{3,21} to obtain more general restrictions for the Higgs
masses in supersymmetric theories. On the other hand such a general
consideration allows one to explain observable CP
violating effects only by
spontaneous CP- violation  \cite{3}. For such a models the supersymmetry
breaking term is the following:
\begin{eqnarray}
\nonumber
V_{soft} & = & m_{1}^{2}|H_{1}|^{2}  + m_{2}^{2}|H_{2}|^{2} +
           m_{3}^{2}|N|^{2} + \\ & + &
            (m_{4}NH_{1}H_{2} +
           \frac{m_{5}}{3}N^{3} + h.c.) +...                       % (4)
\end{eqnarray}
where we omit terms connected with sfermion and gaugino fields and write
only terms connected with Higgs  fields. It is necessary to note that
unlike of the ordinary case (3) in (4) the supersymmetry breaking terms
are not universal.
We want to stress that   usually  it is assumed that supersymmetry
breaking terms given by (3) or (4) for the MSSM and NMSSM
respectively have the given  form for unification energies
$M_{G}\sim (10^{16}-10^{18})GeV$. In MSSM below this scale due to
different
interactions of  Higgs doublets with fermions the mass terms
of $H_{1}$ and  $H_{2}$  differ  from each other.
For the considering case, although the values of the
supersymmetry breaking parameters are changing, when, evaluating from
unification energies to the $M_{s}$,
the mass terms for Higgs doublet are different from the beginning and the
form of supersymmetry breaking terms (4) remains the same.
This means that we can consider
that this form is valid for the supersymmetry breaking scale also.

{}From (1), (2), (4) one obtains the scalar potential for
the Higgs fields $H_{1}$, $H_{2}$, $N$:
\begin{eqnarray}
\nonumber
V & = & \frac{g_{2}^{2}+g_{1}^{2}}{8}((H_{1}^{+}H_{1})^{2}
   + (H_{2}^{+}H_{2})^{2}) + \frac{g_{2}^{2}-g_{1}^{2}}{4}|H_{1}|^{2}
   |H_{2}|^{2}+
(\lambda_{2}^{2}-\frac{1}{2}g_{2}^{2})|H_{1}H_{2}|^{2}  \\
& +  & \lambda_{2}^{2}|N|^{2}(|H_{1}|^{2} + |H_{2}|^{2}) +         %(5)
\lambda_{1}\lambda_{2}((N^{2})^{+}H_{1}H_{2} + h.c.) +
\lambda_{1}^{2}|N|^{4} +    \\
\nonumber
& + & m_{4}(NH_{1}H_{2} + h.c.)
    + \frac{m_{5}}{3}(N^{3} + h.c) +
m_{1}^{2}|H_{1}|^{2} + m_{2}^{2}|H_{2}|^{2} + m_{3}^{2}|N|^{2}
\end{eqnarray}

The potential (5) is positively defined for large values of the Higgs
fields $H_{1}$, $H_{2}$, $N$ vacuum expectation values (VEV) and there
is no problem with vacuum stability in (5).

In general, the supersymmetry and electroweak symmetry breaking
scales are different. This means that the scalar potential at electroweak
scale ($\sim 100GeV$) has the form, different from (5). It is given by
\begin{eqnarray}
\nonumber
V & = & \frac{1}{2}a_{1}(H_{1}^{+}H_{1})^{2} +
\frac{1}{2}a_{2}(H_{2}^{+}H_{2})^{2}  +
a_{3}|H_{1}|^{2}|H_{2}|^{2}+ a_{4}|H_{1}H_{2}|^{2} +
a_{5}|N|^{2}|H_{1}|^{2}   \\
& + & a_{6}|N|^{2}|H_{2}|^{2} +
a_{7}((N^{2})^{+}H_{1}H_{2} + h.c.) + a_{8}|N|^{4} +    %(6)
+m_{4}(NH_{1}H_{2}   \\
\nonumber
& + & h.c.)   + \frac{m_{5}}{3}(N^{3} + h.c) +
m_{1}^{2}|H_{1}|^{2} +
m_{2}^{2}|H_{2}|^{2} + m_{3}^{2}|N|^{2}
\end{eqnarray}
It is easily to see that the potential (6) contains the same terms as (5).
The difference is only  that the coupling constants
in (6) are not connected by  the relations which come from supersymmetry.
The behavior with the energy of the coupling constants $a_{i} (i=1,2,...,8)$
(and also for the Yukawa and gauge couplings) is given by
renormalization group equations. The parameters of potentials (5) and (6)
are connected by the relations:
\begin{eqnarray}
\nonumber
a_{5} & = & a_{6} = \lambda_{2}^{2}, \hspace{0.3cm}
a_{7} = \lambda_{1}\lambda_{2},  \hspace{0.3cm}
a_{8} = \lambda_{1}^{2}   \\
a_{1} & = & a_{2} = \frac{1}{4}(g_{1}^{2}+g_{2}^{2}),  \hspace{0.2cm}
a_{3} = \frac{1}{4}(g_{2}^{2}-g_{1}^{2}),  \hspace{0.2cm}
a_{4} = \lambda_{2}^{2}-\frac{1}{2}g_{2}^{2}     %(7)
\end{eqnarray}
The relations (7) are valid at the supersymmetry breaking scale $M_{s}$
and higher.

4.In our model the electroweak symmetry breaking takes place, when
the Higgs fields $H_{1}$, $H_{2}$, $N$ in potential (6) develop
nonzero VEV's. To provide the electric charge conservation we
shall chose these VEV's in the following form:
\begin{equation}
<H_{1}> =
\left(
\begin{array}{c}
0 \\
v_{1}
\end{array} \right) \\
\hspace{0.5cm}
<H_{2}> =                                                 %(8)
\left(
\begin{array}{c}
v_{2}e^{i\varphi} \\
0
\end{array} \right)
\hspace{0.5cm}
<N> = v_{3}e^{i\alpha}
\end{equation}
For Higgs VEV's given by (9) the potential (7) is given by:
\begin{eqnarray}
\nonumber
V_{o}(v_{1},v_{2},v_{3}) & = &
\frac{1}{2}a_{1}v_{1}^{4} + \frac{1}{2}a_{2}v_{2}^{4} +
         (a_{3}+a_{4})v_{1}^{2}v_{2}^{2} + a_{5}v_{1}^{2}v_{3}^{2} +
          a_{6}v_{2}^{2}v_{3}^{2} \\
& + & 2a_{7}v_{1}v_{2}v_{3}^{2}cos(\varphi-
2\alpha) +
2m_{4}v_{1}v_{2}v_{3}cos(\varphi+\alpha)  + \\
\nonumber
& + & \frac{2}{3}m_{5}v_{3}^{3}cos3\alpha + a_{8}v_{3}^{4} +   %(9)
m_{1}^{2}v_{1}^{2} +
m_{2}^{2}v_{2}^{2} + m_{3}^{2}v_{3}^{2}
\end{eqnarray}
It is easily to see that the potential (9) is invariant for discrete
transformations of the type
\begin{displaymath}
\varphi \rightarrow \varphi + \frac{\pi}{3}, \hspace{0.3cm}
\alpha  \rightarrow \alpha  + \frac{2\pi}{3}, \hspace{0.3cm}
m_{4}  \rightarrow -m_{4},\hspace{0.3cm}
a_{7}   \rightarrow -a_{7}
\end{displaymath}
This means that we can choose (without loss of generality)
\begin{displaymath}
v_{1},v_{2},v_{3}>0;\hspace{0.3cm}
-\frac{\pi}{2} \leq \varphi \leq \frac{\pi}{2}; \hspace{0.3cm}
-\frac{\pi}{6} \leq \alpha \leq \frac{\pi}{6}
\end{displaymath}
For the potential to be a stable one
(this means that for the large values of $v_{i}, i=1.2.3$ it must be
positive) the following conditions must be satisfied:
\begin{eqnarray}
\nonumber
a_{1} > 0,
\hspace {1cm}
a_{2} > 0,
\hspace {1cm}
a_{8} > 0  \\
a_{3} + a_{4} + \sqrt{a_{1}a_{2}} > 0  \\
\nonumber
a_{6} + \sqrt{2a_{1}a_{8}} > 0,                        % (10)
\hspace{0.5cm}
a_{6} + \sqrt{2a_{2}a_{8}} > 0
\end{eqnarray}

The equations for the potential minimum have the form
\begin{eqnarray}
\nonumber
\frac{\partial V}{\partial \varphi} =
\frac{\partial V}{\partial \alpha} = 0  \\
\frac{\partial V}{\partial v_{1}} = \frac{\partial V}{\partial v_{2}}
= \frac{\partial V}{\partial v_{3}} = 0   % (11)
\end{eqnarray}
The first two conditions can be written in the form:
\begin{equation}
m_{4}sin(\varphi+\alpha)
+ a_{7}v_{3}sin(\varphi-2\alpha) = 0 \\   %(12)
\end{equation}
\begin{equation}
2a_{7}v_{1}v_{2}v_{3}sin(2\alpha-\varphi) +
m_{4}v_{1}v_{2}sin(\alpha+\varphi) +
m_{5}v_{3}^{2}sin3\alpha = 0         %  (13)
\end{equation}
The equations (12), (13) have two type of solutions. First, with
$\alpha = \varphi = 0$ corresponds to the case of absence of
spontaneous CP-violation. The second type of solution with nonzero
phases $\alpha, \varphi$ corresponds to the case of spontaneous CP-
violation.  For equations (12), (13) to have nonzero solutions for
$\varphi$ and $\alpha$ the following condition must be satisfied:
\begin{equation}
3a_{7}v_{1}v_{2}cos\varphi + \frac{v_{3}m_{5}}{m_{4}}    % (14)
(a_{7}v_{3}cos2\alpha + m_{4}cos\alpha) = 0
\end{equation}
Three last conditions of (11)
can be written in the
following form:
\begin{eqnarray}
\nonumber
&a_{1}v_{1}^{3}& + (a_{3} + a_{4})v_{1}v_{2}^{2} + a_{5}v_{1}v_{3}^{2}
 +  m_{1}^{2}v_{1}  \\
& + &(m_{4}cos(\varphi + \alpha) + a_{7}v_{3}cos(\varphi-
2\alpha))v_{3}v_{2}
 = 0                      % (15)
\end{eqnarray}
\begin{eqnarray}
\nonumber
&a_{2}v_{2}^{3}& + (a_{3} + a_{4})v_{2}v_{1}^{2} + a_{6}v_{2}v_{3}^{2} +
m_{2}^{2}v_{2} +  \\
&+&(m_{4}cos(\varphi + \alpha) + a_{7}v_{3}cos(\varphi - 2\alpha))v_{3}v_{1}
= 0                        % (16)
\end{eqnarray}
\begin{eqnarray}
\nonumber
2a_{8}v_{3}^{3} + (a_{5}v_{1}^{2} + a_{6}v_{2}^{2} +
2a_{7}v_{1}v_{2}cos(\varphi - 2\alpha))v_{3}   \\
+ m_{5}v_{3}^{2}cos3\alpha +
m_{4}v_{1}v_{2}cos(\varphi + \alpha) + m_{3}^{2}v_{3} = 0  % (17)
\end{eqnarray}
The equations (15)-(17) are satisfied for $v_{1} = v_{2} = v_{3} = 0$.
We are looking for solutions with
\begin{equation}
v_{1}\neq0, v_{2}\neq0, v_{3}\neq0.                   %  (18)
\end{equation}
For the solutions (18) to be a real minimum of the potential
the value of $V_{o}(v_{1},v_{2},v_{3})$
(where $v_{1}, v_{2}, v_{3}$ are nonzero solutions of equations
(12), (13), (15)-(17)) must be negative:
\begin{eqnarray}
\nonumber
V(v_{1}, v_{2}, v_{3}) & = &
\frac{1}{2}(\frac{1}{3}m_{5}
v_{3}^{3}cos3\alpha + m_{4}v_{1}v_{2}v_{3}cos(\varphi + \alpha)  %(19)
+ m_{1}^{2}v_{1}^{2}   \\
& + & m_{2}^{2}v_{2}^{2} + m_{3}^{2}v_{3}^{2})
< V(0, 0, 0) = 0
\end{eqnarray}
We must take into account also the following: for the solution with
nonzero values of $\alpha, \varphi$ and spontaneous CP- violation
to be the physical minimum of the potential the potential value
in this point must be less than its value for $\alpha = 0, \varphi = 0$.
This means that the following condition must be satisfied:
\begin{equation}
m_{5}sin\frac{\varphi + \alpha}{2} sin\frac{\varphi -2\alpha}{2} < 0   %(20)
\end{equation}

5.Let us consider now the masses of Higgs particles. After the spontaneous
breaking of electroweak symmetry it appears five neutral and one complex
charged fields. The goldstone mode for neutral fields can be excluded
by the transformation:
\begin{eqnarray}
\nonumber
\xi_{1}^{o} & = & v_{1} + \frac{1}{\sqrt{2}}(\Phi_{1}cos\beta -
\Phi_{2}sin\beta - i(Gcos\beta - Asin\beta))  \\                 %(21)
\xi_{2}^{o} & = & e^{i\varphi}(v_{2} + \frac{1}{\sqrt{2}}(\Phi_{1}sin\beta +
\Phi_{2}cos\beta + i(Gsin\beta + Acos\beta))) \\
\nonumber
N & = & e^{i\alpha}(N_{1} + iN_{2})
\end{eqnarray}
where $\Phi_{1}, \Phi_{2}, A, N_{1}, N_{2}$ are new neutral fields,
G is the goldstone mode, $tan\beta = \frac{v_{2}}{v_{1}}$.
The $5 \times 5$ symmetric mass matrix $M_{ij}^{2}, i,j=1,...,5$
for fields
$\Phi_{1}, \Phi_{2}, A, N_{1}, N_{2}$  for the case of the presence of
CP- violation is given by formula (A1) of the Appendix A.
For the case of absence of CP- violation the mass matrix has more
simple form: it consists of two matrices,
one of them $3\times3$ is the mass matrix of CP- even fields
$\Phi_{1}, \Phi_{2}, N_{1}$, the second one is $2\times2$ mass matrix of
CP- odd fields $A, N_{2}$. They are given by formulae (A2) and (A3) of
the Appendix A.
For the charged Higgs particles one can exclude the Goldstone mode
by the transformation:
\begin{eqnarray}
\nonumber
\xi_{1}^{+} = -G^{+}cos\beta + e^{i\varphi}H^{+}sin\beta  \\       %(22)
\xi_{2}^{-} = e^{i\varphi}G^{-}sin\beta + H^{-}cos\beta
\end{eqnarray}
The mass of charged Higgs is given by:
\begin{equation}
m^{2}_{H^{+}} = - \frac{2(m_{4}cos(\varphi + \alpha) + a_{7}v_{3}cos(
\varphi - 2\alpha))v_{3}}{sin2\beta} - a_{4}\eta^{2}                  % (23)
\end{equation}
where
\begin{equation}
\eta^{2} = v_{1}^{2} + v_{2}^{2} = (174GeV)^{2}                      % (24)
\end{equation}
It is necessary to note that the condition of positiveness of
charged Higgs mass squared provides the conservation of electric
charge i.e. the Higgs boson VEV's have the form (8) \cite{9}

6.To obtain the bounds for masses of the Higgs particles
one must  investigate the possible restrictions for parameters
of Higgs potential (6).
The restrictions for the coupling constants $a_{i}, i=1,...8$
one can obtain, by
analyzing the renormalization group equations
for $a_{i}, i=1,...8$, gauge $g_{r}, r=1,2,3$ and t and b
quark Yukawa couplings
$h_{t}=\frac{m_{t}}{v_{2}}$, $h_{b}=\frac{m_{b}}{v_{1}}$
(one can neglect other Yukawa couplings because of their smallness with
regard to $h_{t}$ and $h_{b}$).

In the region between
the electroweak symmetry breaking scale and supersymmetry breaking scale they
have the form \cite{6,22} ($t=ln(\frac{Q}{100GeV}$):
\begin{eqnarray}
\nonumber
16\pi^{2}\frac{\partial a_{i}}{\partial t} & = & f_{i}, i=1,...8  \\
\nonumber
16\pi^{2}\frac{\partial h_{t}}{\partial t} & = &  h_{t}(\frac{9}{2}h_{t}^{2}
+\frac{1}{2}h_{b}^{2} -
\frac{17}{12}g_{1}^{2} - \frac{9}{4}g_{2}^{2} - 8g_{3}^{2})  \\  % (25)
16\pi^{2}\frac{\partial h_{b}}{\partial t} & = &  h_{b}(\frac{9}{2}h_{b}^{2}
+\frac{1}{2}h_{t}^{2} -
\frac{5}{12}g_{1}^{2} - \frac{9}{4}g_{2}^{2} - 8g_{3}^{2})  \\
\nonumber
16\pi^{2}\frac{\partial g_{i}}{\partial t} & = & -c_{i}g_{i}^{3}
\end{eqnarray}
where
\begin{displaymath}
c_{1} = -7,
\hspace{0.5cm}
c_{2} = 3,
\hspace{0.5cm}
c_{3} = 7
\end{displaymath}
and $f_{i}$, i=1,...,8 are given by the formula (B1) of the Appendix B.
The values of gauge constants $g_{r}$ (r=1,2,3) at electroweak breaking scale
$\sim M_{Z}$  are the following \cite{13}:
\begin{displaymath}
g_{1} = 0.358
\hspace{0.4cm}
g_{2} = 0.651
\hspace{0.4cm}
g_{3} = 1.218
\end{displaymath}
Above the supersymmetry breaking scale some of the
constants $a_{i}$ (i=1,...8)
are unified according to equations (7), and
the renormalization group equations should be written for the
quantities
$\lambda_{1}$, $\lambda_{2}$, $h_{t}$, $h_{b}$, $g_{r}$
instead of the quantities
$a_{i}$, $h_{t}$, $h_{b}$, $g_{r}$ (r=1,2,3; i=1,...8).
These equations are as follows \cite{23} $(t=ln\frac{Q}{M_{s}})$:
\begin{eqnarray}
\nonumber
16\pi^{2}\frac{dh_{t}}{dt} & = & h_{t}(6h_{t}^{2} + h_{b}^{2} +
\lambda_{2}^{2} -
\frac{13}{9}g_{1}^{2} - 3g_{2}^{2} - \frac{16}{3}g_{3}^{2})  \\
\nonumber
16\pi^{2}\frac{dh_{b}}{dt} & = & h_{b}(6h_{b}^{2} + h_{t}^{2} +
\lambda_{2}^{2} -
\frac{10}{9}g_{1}^{2} - 3g_{2}^{2} - \frac{16}{3}g_{3}^{2})  \\
16\pi^{2}\frac{d\lambda_{2}}{dt} & = & \lambda_{2}(2\lambda_{1}^{2} +
4\lambda_{2}^{2} + 3(h_{t}^{2} + h_{b}^{2}) - g_{1}^{2} -3g_{2}^{2})  \\
\nonumber
 16\pi^{2}\frac{d\lambda_{1}}{dt} & = & 6\lambda_{1}(\lambda_{1}^{2} + %(26)
\lambda_{2}^{2}) \\
\nonumber
16\pi^{2}\frac{dg_{i}}{dt} & = & -c_{i}g_{i}^{3};
\hspace{0.5cm}
c_{1} = -11,
\hspace{0.2cm}
c_{1} = -1,
\hspace{0.2cm}
c_{3} = 3
\end{eqnarray}
To obtain the restrictions for $a_{i}$ at scale $Q=M_{Z}$
we investigate numerically the differential equations (25) from $Q=M_{Z}$
to $Q=M_{s}$ and (26) from $Q=M_{s}$ to $Q=M_{G}$.
These restrictions for $a_{i}$ arise in the following way. We require
that in the region from $M_{Z}$ to $M_{s}$ and from $M_{s}$ to $M_{G}$
all of the coupling constants must be small so that the perturbation
theory applies.
This will give some restrictions for values of scalar coupling constants
$a_{i}$ at $Q=M_{Z}$ and so for the Higgs masses. One must take into account
also the vacuum stability condition which will give an additional
restrictions for $a_{i}$.

More precisely, the
perturbation theory to be applied, the
quantities $a_{i} (i=1,...8)$, $h_{t}$, $h_{b}$ at $Q=M_{Z}$
must satisfy the following conditions:
\begin{equation}
\frac{h_{t}^{2}}{4\pi}, \hspace{0.2cm}
\frac{h_{b}^{2}}{4\pi}, \hspace{0.2cm}
\frac{|a_{i}|}{4\pi}   < 1                      % (27)
\end{equation}
The conditions (27) must be valid for also for
$M_{Z}<Q<M_{s}$.
For $M_{s}<Q<M_{G}$, conditions (27) are replaced by the conditions
(k=1,2):
\begin{equation}
\frac{\lambda_{k}^{2}}{4\pi}, \hspace{0.2cm}
\frac{h_{b}^{2}}{4\pi}, \hspace{0.2cm}
\frac{h_{t}^{2}}{4\pi}   < 1                      % (28)
\end{equation}
Actually, the conditions (27), (28) are not enough to provide the
perturbation theory validity. Severely speaking, the condition
of smallness of derivative of coupling constants must be satisfied:
\begin{equation}
\frac{\partial a_{i}}{\partial t}, \hspace{0.2cm}
\frac{\partial \lambda_{k}}{\partial t}, \hspace{0.2cm}        %(29)
\frac{\partial h_{t}}{\partial t}, \hspace{0.2cm}
\frac{\partial h_{b}}{\partial t}  <  A
\end{equation}
where A is a number $\sim 1$.

Let us proceed to concrete results. We will consider that the supersymmetry
breaking scale $M_{s}$ is between 100GeV and 10000GeV and the
unification scale is between $10^{16}GeV$ and $10^{18}GeV$.
The values of t and b quark masses at scale $M_{Z}$ we take equal to
$m_{t}=(175\pm15)$GeV, $m_{b}=(3.5\pm0.5)$GeV
\cite{24,25}.

We  investigate numerically the renormalization group equations, taking
into account the condition of the validity of the perturbation theory
and the vacuum stability condition.
If the supersymmetry breaking scale is
$M_{s}=M_{Z}$,
one obtains the restrictions for $\lambda_{1}, \lambda_{2},
h_{t}$ and $h_{b}$:
\begin{equation}
\nonumber
|\lambda_{1}|\leq 0.5, \hspace{0.2cm}
|\lambda_{2}|\leq 0.6, \hspace{0.2cm}
1.0 \leq |h_{t}| \leq 1.1, \hspace{0.2cm}               \\           % (30)
|h_{b}|\leq 1.0
\end{equation}
For $M_{s}=1000GeV$ and $M_{s}=1000GeV$ the restrictions for
parameters $a_{i}(i=1,...,8), h_{t}$ and $h_{b}$ are given in Appendix B.

The Higgs masses depend also on mass parameters of the
potential and
Higgs boson  VEV's $v_{1},v_{2},v_{3}$.
The mass parameters of this potential
$m_{4}$, $m_{5}$,
$m_{1}^{2}$, $m^{2}_{2}$,  $m_{3}^{2}$
are connected with supersymmetry breaking and are of order of supersymmetry
breaking scale.
The Higgs VEV's $v_{1}, v_{2}$ are restricted by the condition
(24).
The restrictions for their ratio
\begin{equation}
\frac{v_{2}}{v_{1}} = \frac{v_{2}}{\sqrt{\eta^{2} - v_{2}^{2}}} =      %{31)
tan\beta
\end{equation}
are obtained from values of t and b-quark masses.
{}From (31) one obtains for
$m_{t}=(175\pm15)$GeV, $m_{b}=(3.5\pm0.5)$GeV
\begin{equation}
1 < tan\beta < 60               % (32)
\end{equation}
The singlet's VEV $v_{3}$, generally speaking is not connected with the
supersymmetry breaking and so its scale is not coincide with the
supersymmetry breaking scale $M_{s}$.
However the equations of the potential minimum
(15)-(17) give an additional restrictions
for this parameter.

Let as try to analyze qualitatively which bounds one can obtain
for neutral and charged Higgs masses, when apply
the above obtained restrictions.

First of all let us consider the case of absence of spontaneous
CP-breaking.

It is easily to see from the formula (23)
and the previous analysis that the charged
Higgs particle mass can be as heavy as supersymmetry breaking scale.

As concerned the neutral Higgs masses, as we have mentioned above,
the mass matrix consists of two separate $3\times 3$ and $2\times 2$
matrices for CP even and CP odd particles respectively. The analysis
of these mass matrix shows that the upper bound of mass of one CP even
particle is of order $\eta=174GeV$ and four other particles masses,
generally speaking, are of order of supersymmetry breaking scale $M_{s}$.
This follows, in particular, from the fact that determinant of CP even
particles mass matrix is proportional to $\eta^{2}M_{s}^{4}$ and
determinant of CP odd particle mass matrix is proportional to $M_{s}^{4}$.

The case of the presence of spontaneous CP-violation is more
complicated: here one has $5\times 5$ neutral Higgses  mass matrix.
In this case the conditions of positiveness of Higgs boson masses are very
important. These conditions are equivalent to the positiveness
of determinant and diagonal minors of matrix (A1). The analytical
investigations of these conditions (taking into account also the
condition of positiveness of charged Higgs particle) gives the
following result: for supersymmetry breaking scale $M_{s} \sim 100GeV$
these conditions can't be satisfied simultaneously, i.e for
$M_{s} \sim 100GeV$ the spontaneous CP breaking can't take place.
As concerned $M_{s} \sim 1000GeV$ and $M_{s} \sim 10000GeV$,
the condition of the neutral Higgs mass matrix to be positively
determined gives the following inequality which must be satisfied
for energies $\sim 100GeV$:
\begin{eqnarray}
\nonumber
max((a_{3} + a_{4} - \sqrt{a_{1}a_{2}})\frac{\eta^{2}sin2\beta}{2v_{3}^{2}}
,\frac{a_{7}^{2}\eta^{2}sin2\beta}{2a_{8}v_{3}^{2}})<       \\
< -a_{7}\frac{sin3\alpha}{sin(\varphi+\alpha)} < (a_{3} + a_{4} +  %(33)
\sqrt{a_{1}a_{2}}) \frac{\eta^{2} sin2\beta}{2v_{3}^{2}}
\end{eqnarray}
This condition is obtained from requirement that the fourth order
diagonal two minors of matrix (A1) which can be obtained, when
first or fourth lines and columns are excluded must be positive.
Let us note also that for $M_{s} \sim 100GeV$ the inequalities (33)
become the equalities due to the relations (7) and it is not possible to
provide the positiveness of the diagonal minors, as we mentioned  above.
The conditions (33) can be satisfied only due to the fact that
the coupling constants $a_{i}$, i=1,...,8, for $M_{s} \sim 1000GeV$ or
$M_{s} \sim 10000GeV$ are not satisfying relations (7) for
energies $\sim 100GeV$ due to the their evolution from supersymmetry
breaking scale to electroweak breaking scale according
renormalization group equations (25).

Using the condition (33), we come to the following restrictions for
the neutral Higgs particles masses in the case of presence of CP-
violation. The
lightest neutral Higgs has mass of order of $\eta$ multiplied by the
factor of order of radiative corrections to the coupling constants $a_{i}$,
i=1,...,8, when, evaluating from 1000GeV or 10000GeV to 100GeV,
i.e., in general, sufficiently small. Two other neutral Higgs boson have
masses of order of $\sim \eta$ and masses of two last neutral Higgses can
be as heavy as supersymmetry breaking scale $M_{s}$. These results are
obtained, in particular, from the fact that due to the condition (33)
determinant of matrix (A1) is proportional to $M_{s}^{4}\eta_{6}$
multiplied by the factor(s) of order of radiative corrections
to the coupling constants $a_{i}$.

It is interesting to analyze the case, when one or two neutral Higgs
particles are almost $SU(2) \times U(1)$ singlets. In this case the situation
is the following: if both of singlets are much heavier than electroweak
breaking scale, then, as it follows from the previous analysis, the remaining
neutral scalar particles have a masses of order of $\eta$ or smaller.
The following statement is also true: if one of the
neutral Higgs boson is much heavier than $\eta$, then it is almost
singlet. This means that in all cases one has three neutral
visible (nonsinglet) particles with "small" ($\leq \eta$) masses.

Using the condition (33) and equation (12) one obtains from (23) the
following restriction for the mass of charged Higgs particle:
\begin{equation}
m_{H^{+}}^{2} < (a_{3} + \sqrt{a_{1}a_{2}})\eta^{2}         %(34)
\end{equation}
This means that in the case of the presence of spontaneous CP-violation
mass of charged Higgs particle is also of order of $\eta$.

We want to stress that the results, we obtained, in the case of the presence
of CP-violation are very similar to results obtained in \cite{14}
for the case of nonsupersymmetric two Higgs doublet model: for
both of
cases three neutral and one charged Higgses have masses of order of $\eta$
or less.

7.Let us proceed to numerical investigations of bounds of
Higgs boson masses. The Higgs boson masses depend on the parameters
$a_{i}$, i=1,...,8; $m_{k}^{2}$, k=1,...,5; $v_{3}$; $tan\beta$;
$\varphi$; $\alpha$ (for the case of the presence of CP-violation);
$\eta=174GeV$. The restrictions for $a_{i}$, i=1,...,8
are given by formulae (10), (30), (B2), (B3). The parameters
$m_{k}^{2}$, k=1,...,5 are connected with the supersymmetry breaking
and we take them $\sim M_{s}$ or smaller. Besides this, these
parameters,
as well as $v_{3}$ and $\varphi$, $\alpha$ are restricted by the
conditions  (12)-(17),(19),(20) and (33) for the case of the presence of
spontaneous CP-violation and also by other conditions connected with
positiveness of Higgs boson masses squared. The restrictions for $tan\beta$
is given by formula (32).

All of the restrictions mentioned above give the bounds for the
Higgs boson
masses which have been presented in Fig. 1 - 5.  These results have been
obtained with accuracy of order of $10\%$.

As we have mentioned above,
in the case of absence of CP- violation all Higgs particles besides the
lightest one can be as heavy, as supersymmetry breaking scale $M_{s}$.
The upper bound for the lightest neutral CP- even Higgs mass
($m_{h_{1}}$)
as a function
of $tan\beta$ for the cases
$M_{s} \sim$ 100GeV, 1000GeV and 10000GeV are presented in Fig. 1.
As we can see, in this case $m_{h_{1}} \leq$ 105GeV, 135GeV and 150GeV
respectively for $M_{s} \sim$ 100GeV, 1000GeV and 10000GeV.

In the case of the presence of CP- violation only two neutral Higgses can
be as heavy as supersymmetry breaking scale $M_{s}$. The upper bounds as a
functions of $tan\beta$ we obtained for masses of remaining neutral and
charged Higgses ($m_{h_{1}}$, $m_{h_{2}}$, $m_{h_{3}}$ and  $m_{H^{+}}$
respectively) are presented in Fig. 2-5.
As it follows from Fig.2-5,
\begin{eqnarray}
\nonumber
m_{h_{1}} \leq 30GeV, \hspace{2cm}
m_{h_{2}} \leq 95GeV,      \\                          %(35)
m_{h_{3}} \leq 135GeV, \hspace{2cm}
m_{H^{+}} \leq 115GeV,
\end{eqnarray}
for $M_{s} \sim 1000GeV$ and
\begin{eqnarray}
\nonumber
m_{h_{1}} \leq 35GeV, \hspace{2cm}
m_{h_{2}} \leq 100GeV,    \\
m_{h_{3}} \leq 150GeV, \hspace{2cm}                      %(36)
m_{H^{+}} \leq 145GeV,
\end{eqnarray}
for $M_{s} \sim 10000GeV$.

The experimental restrictions for Higgs boson masses
in supersymmetric standard model Higgses are the following \cite{26}:
\begin{equation}
m_{h_{1}} > 45GeV, \hspace{2cm}        % (37)
m_{H^{+}} > 45GeV
\end{equation}
There is an obvious contradiction between obtained bound (36) for
$m_{h_{1}}$ and the experimental restriction (37). But it is necessary
to note that the NMSSM includes additional $SU(2) \times U(1)$ singlets
in Higgs sector in compare with the MSSM and so the experimental bounds
(37) in general are not true for NMSSM neutral Higgs sector. More
precisely, to avoid the contradiction with experiment the lightest
neutral Higgs $h_{1}$ must be almost singlet to escape the detection.
The preliminary analysis shows that there exists a region of parameter
space where such a situation can take place.
As we have mentioned above in this case
three detectable (i.e. nonsinglet) particles with mass of order of $\eta$
or smaller is always exist.

8.Thus we have investigated bounds for Higgs boson masses in NMSSM
which arise from renormalization group equations and scalar
Higgs potential analysis. For the case of absence of spontaneous
CP- violation we obtain that neutral CP- even Higgs particle with mass
smaller
than $\eta$ exists. The above restriction is very close to one obtained
in MSSM.
We have shown that if $M_{s} \sim 100GeV$, spontaneous
CP- violations isn't possible. If supersymmetry is broken at higher
scales, then the scenario with spontaneous CP- violation to
be real the lightest neutral Higgs must be almost $SU(2) \times U(1)$
singlet . In this case charged Higgs boson have mass smaller than
$\eta$ and at least three detectable (i.e. nonsinglet) neutral Higgs
boson exist with masses of order of $\eta$. Thus the restrictions obtained
for the case of the presence of spontaneous CP- violation are much stronger
than in MSSM.

So our analysis of the problem of Higgs masses in NMSSM
with the spontaneous CP-breaking shows that the considering model
leads to the predictions for Higgs particles masses which can be
verified experimentally in the near future. This model can be
considered as an interesting alternative of MSSM.
The investigation of properties  of the considered model
in more detail is required. In particular, the correlation between
Higgs masses and decay rates in the case, when some of the neutral
Higgses are nearly singlets must be investigated.
This work is in progress now.

   The research described in this publication was made possible in part
by Grant N MVU000 from the International Science Foundation.

\vspace{1cm}
\renewcommand{\theequation}{A.\arabic{equation}}
\setcounter{equation}0
\appendix
\begin{center}
    APPENDIX A
\end{center}
\vspace{1cm}

The $5\times5$  symmetric mass matrix $M_{ij}^{2}$ (i,j = 1,...,5) of
neutral Higgs particles $\Phi_{1}, \Phi_{2}, A, N_{1}, N_{2}$
for the case of the presence CP- violation is given by:
\begin{eqnarray}
\nonumber
M_{11}^{2} & = &
(\frac{1}{2}(a_{1} + a_{2} - 2(a_{3} + a_{4}))cos^{2} 2\beta -
(a_{2} - a_{1})cos2\beta + (a_{1} + a_{2} + 2(a_{3} + a_{4})))\eta^{2}  \\
\nonumber
M_{12}^{2} & = &
(-\frac{1}{2}(a_{1} + a_{2} - 2(a_{3} + a_{4}))cos2\beta sin2\beta +
\frac{1}{2}(a_{2} - a_{1})sin2\beta )\eta^{2}  \\
\nonumber
M_{13}^{2} & = & 0  \\
\nonumber
M_{14}^{2} & = &
(a_{5} + a_{6} + (a_{5} - a_{6})cos2\beta + a_{7} sin2\beta
(cos(\varphi - 2\alpha) + \frac{sin3\alpha}{sin(\varphi + \alpha)}))
v_{3} \eta  \\
\nonumber
M_{15}^{2} & = & 3a_{7}v_{3} \eta sin2\beta sin(\varphi - 2\alpha)  \\
\nonumber
M_{22}^{2} & = &
\frac{1}{2}(a_{1} + a_{2} - 2(a_{3} + a_{4})) \eta^{2} sin^{2} 2\beta
- \frac{2a_{7}v_{3}^{2} sin3\alpha}{sin2\beta sin(\varphi + \alpha)}    \\
M_{23}^{2} & = &                                                      %(A1)
0  \\
\nonumber
M_{24}^{2} & = &
((a_{6} - a_{5})sin2\beta + a_{7} cos2\beta
(cos(\varphi - 2\alpha) + \frac{sin3\alpha}{sin(\varphi + \alpha)}))
v_{3} \eta  \\
\nonumber
M_{25}^{2} & = & 3a_{7}v_{3} \eta cos2\beta sin(\varphi - 2\alpha)  \\
\nonumber
M_{33}^{2} & = &
- \frac{2a_{7}v_{3}^{2} sin3\alpha}{sin2\beta sin(\varphi + \alpha)}    \\
\nonumber
M_{34} & = & -a_{7}v_{3} \eta sin(\varphi - 2\alpha)  \\
\nonumber
M_{35}^{2} & = & a_{7}v_{3} \eta
(3cos(\varphi - 2\alpha) - \frac{sin3\alpha}{sin(\varphi + \alpha)})  \\
\nonumber
M_{44}^{2} & = &
4a_{8}v_{3}^{2} + \frac{a_{7} \eta^{2} sin2\beta sin(\varphi - 2\alpha)}
{2}(3cot3\alpha + cot(\varphi + \alpha))   \\
\nonumber
M_{45}^{2} & = & -2a_{7} \eta^{2} sin2\beta sin(\varphi - 2\alpha)    \\
\nonumber
M_{55}^{2} & = & -\frac{a_{7} \eta^{2} sin2\beta}{2} (9sin(\varphi -
2\alpha)cot3\alpha + 3cos(\varphi - 2\alpha) +
\frac{sin3\alpha}{sin(\varphi + \alpha)})
\end{eqnarray}
where $\eta^{2} = v_{1}^{2} + v_{2}^{2} = (174GeV)^{2}$.

For the case of the absence of CP- violation the $3\times3$
symmetric mass matrix $(M^{2}_{sc})_{ij}$ (i,j = 1,2,3)
for the scalar $\Phi_{1}, \Phi_{2}, N_{1}$ fields is:
\begin{eqnarray}
\nonumber
(M_{sc}^{2})_{11} & = &
(\frac{1}{2}(a_{1} + a_{2} - 2(a_{3} + a_{4}))cos^{2} 2\beta -
(a_{2} - a_{1})cos2\beta + (a_{1} + a_{2} + 2(a_{3} + a_{4})))\eta^{2}  \\
\nonumber
(M_{sc}^{2})_{12} & = &
(-\frac{1}{2}(a_{1} + a_{2} - 2(a_{3} + a_{4}))cos2\beta sin2\beta +
\frac{1}{2}(a_{2} - a_{1})sin2\beta )\eta^{2}  \\
\nonumber
(M_{sc}^{2})_{13} & = & (a_{5} + a_{6} + (a_{5} - a_{6})cos2\beta) v_{3} \eta
+ (m_{4} + 2a_{7}v_{3}) \eta sin2\beta \\
(M_{sc}^{2})_{22} & = &                                                %(A2)
\frac{1}{2}(a_{1} + a_{2} - 2(a_{3} + a_{4})) \eta^{2} sin^{2} 2\beta
- \frac{2(m_{4} + a_{7}v_{3})v_{3}}{sin2\beta}    \\
\nonumber
(M_{sc}^{2})_{23} & = & (a_{6} - a_{5})sin2\beta v_{3} \eta
- (m_{4} + 2a_{7}v_{3}) \eta cos2\beta \\
\nonumber
(M_{sc}^{2})_{33} & = &
4a_{8}v_{3}^{2} + m_{5}v_{3} - \frac{m_{4} \eta^{2} sin2\beta}
{2v_{3}}
\end{eqnarray}
The $2\times2$ symmetric mass
matrix $(M_{p}^{2})_{ij}$ (i,j = 1,2) for the pseudoscalar fields
$A, N_{2}$  is as follows:
\begin{eqnarray}
\nonumber
(M_{p}^{2})_{11} & = &
- \frac{2(m_{4} + a_{7}v_{3})v_{3}}{sin2\beta}    \\
(M_{p}^{2})_{12} & = &
-(m_{4} - 2a_{7}v_{3}) \eta   \\                               %(A3)
\nonumber
(M_{p}^{2})_{22} & = & - 3m_{5}v_{3} - (m_{4} + 4a_{7}v_{3})
\frac{\eta^{2} sin2\beta}{2v_{3}}
\end{eqnarray}
\vspace{1cm}
\renewcommand{\theequation}{B.\arabic{equation}}
\setcounter{equation}0
\appendix
\begin{center}
    APPENDIX B
\end{center}
\vspace{1cm}
The expressions for $f_{i}$, i=1,...,8 in (27) are given by:
\begin{eqnarray}
\nonumber
f_{1} & = &
12a_{1}^{2}+4a_{3}^{2}+4a_{3}a_{4} + 2a_{4}^{2} + 2a_{5}^{2} -  \\
\nonumber   & & \mbox{}
a_{1}(3g_{1}^{2} +9g_{2}^{2}) +\frac{3}{4}g_{1}^{4} +\frac{9}{4}g_{2}^{4}
+ \frac{3}{2}g_{1}^{2}g_{2}^{2} + 12a_{1}h_{b}^{2} - 12h_{b}^{4}  \\
\nonumber
f_{2} & = &
12a_{2}^{2}+4a_{3}^{2}+4a_{3}a_{4} + 2a_{4}^{2} + 2a_{6}^{2} -  \\
\nonumber  & & \mbox{}
a_{2}(3g_{1}^{2} +9g_{2}^{2}) +\frac{3}{4}g_{1}^{4} +\frac{9}{4}g_{2}^{4}
+ \frac{3}{2}g_{1}^{2}g_{2}^{2}  + 12a_{2}h_{t}^{2} - 12h_{t}^{4}  \\
\nonumber
f_{3} & = &  2(a_{1} + a_{2})(3a_{3} + a_{4}) + 4a_{3}^{2} + 2a_{4}^{2} +
 2a_{5}a_{6} - a_{3}(3g_{1}^{2} +9g_{2}^{2})  \\
\nonumber & & \mbox{}
+\frac{3}{4}g_{1}^{4} +\frac{9}{4}g_{2}^{4}
- \frac{3}{2}g_{1}^{2}g_{2}^{2}  + 6a_{3}(h_{t}^{2} + h_{b}^{2})
- 12 h_{t}^{2} h_{b}^{2}   \\
f_{4} & = & 2a_{4}(a_{1} + a_{2} + 4a_{3} + 2a_{4}) + 4a_{7}^{2} -  \\  % (B1)
\nonumber & & \mbox{}
a_{4}(3g_{1}^{2} +9g_{2}^{2}) + 3g_{1}^{2}g_{2}^{2}  + 6a_{4}(h_{t}^{2} +
h_{b}^{2})  + 12 h_{t}^{2} h_{b}^{2} \\
\nonumber
f_{5} & = & 2a_{5}(3a_{1} + 2a_{5} + 4a_{8}) + 2a_{6}(2a_{3} + a_{4}) + \\
\nonumber  & & \mbox{}
8a_{7}^{2} - \frac{1}{2}a_{5}(3g_{1}^{2} + 9g_{2}^{2}) + 6a_{5}h_{b}^{2}  \\
\nonumber
f_{6} & = & 2a_{6}(3a_{1} + 2a_{6} + 4a_{8}) + 2a_{5}(2a_{3} + a_{4}) +\\
\nonumber  & & \mbox{}
8a_{7}^{2} - \frac{1}{2}a_{6}(3g_{1}^{2} + 9g_{2}^{2}) + 6a_{6}h_{t}^{2}  \\
\nonumber
f_{7} & = &  2a_{7}(a_{3} + 2a_{4} + 2a_{5} + 2a_{6} + 2a_{8})  \\
\nonumber & & \mbox{}
- \frac{1}{2}a_{7}(3g_{1}^{2} + 9g_{2}^{2}) + 3a_{7}(h_{t}^{2} + h_{b}^{2})  \\
\nonumber
f_{8} & = &  2a_{5}^{2} + 2a_{6}^{2} + 4a_{7}^{2} + 20a_{8}^{2}
\end{eqnarray}
For $M_{s}$=1000GeV one obtains the following bound for coupling constants
$a_{i}$, i=1,...,8
\begin{eqnarray}
\nonumber
0.13 \leq a_{1} \leq 0.24  \hspace{0.3cm}
0.23 \leq a_{2} \leq 0.3 \hspace{0.3cm}
0.06 \leq a_{3} \leq 0.19\\
-0.31 \leq a_{4} \leq 0.15 \hspace{0.3cm}
a_{5} \leq 0.36 \hspace{0.3cm} \hspace{0.3cm}                      %(B2)
a_{6} \leq 0.33\\
\nonumber
|a_{7}| \leq 0.19 \hspace{0.3cm}
a_{8} \leq 0.33 \ \hspace{0.3cm}
1 \leq |h_{t}| \leq 1.1  \hspace{0.3cm}
|h_{b}| \leq 1
\end{eqnarray}
And for $M_{s}=10000GeV$:
\begin{eqnarray}
\nonumber
0.12 \leq a_{1} \leq 0.35 \hspace{0.3cm}
0.34 \leq a_{2} \leq 0.35 \hspace{0.3cm}
0.03 \leq a_{3} \leq 0.31\\
-0.43 \leq a_{4} \leq 0.26 \hspace{0.3cm}
a_{5} \leq 0.46 \hspace{0.3cm}              %(B3)
a_{6} \leq 0.40 \hspace{0.3cm} \\
\nonumber
|a_{7}| \leq 0.22 \hspace{0.3cm}
a_{8} \leq 0.3 \hspace{0.3cm}
1.0 \leq |h_{t}| = 1.1  \hspace{0.3cm}
h_{b} \leq 1.1
\end{eqnarray}

\begin{titlepage}
\setcounter{page}{0}
\epsfbox{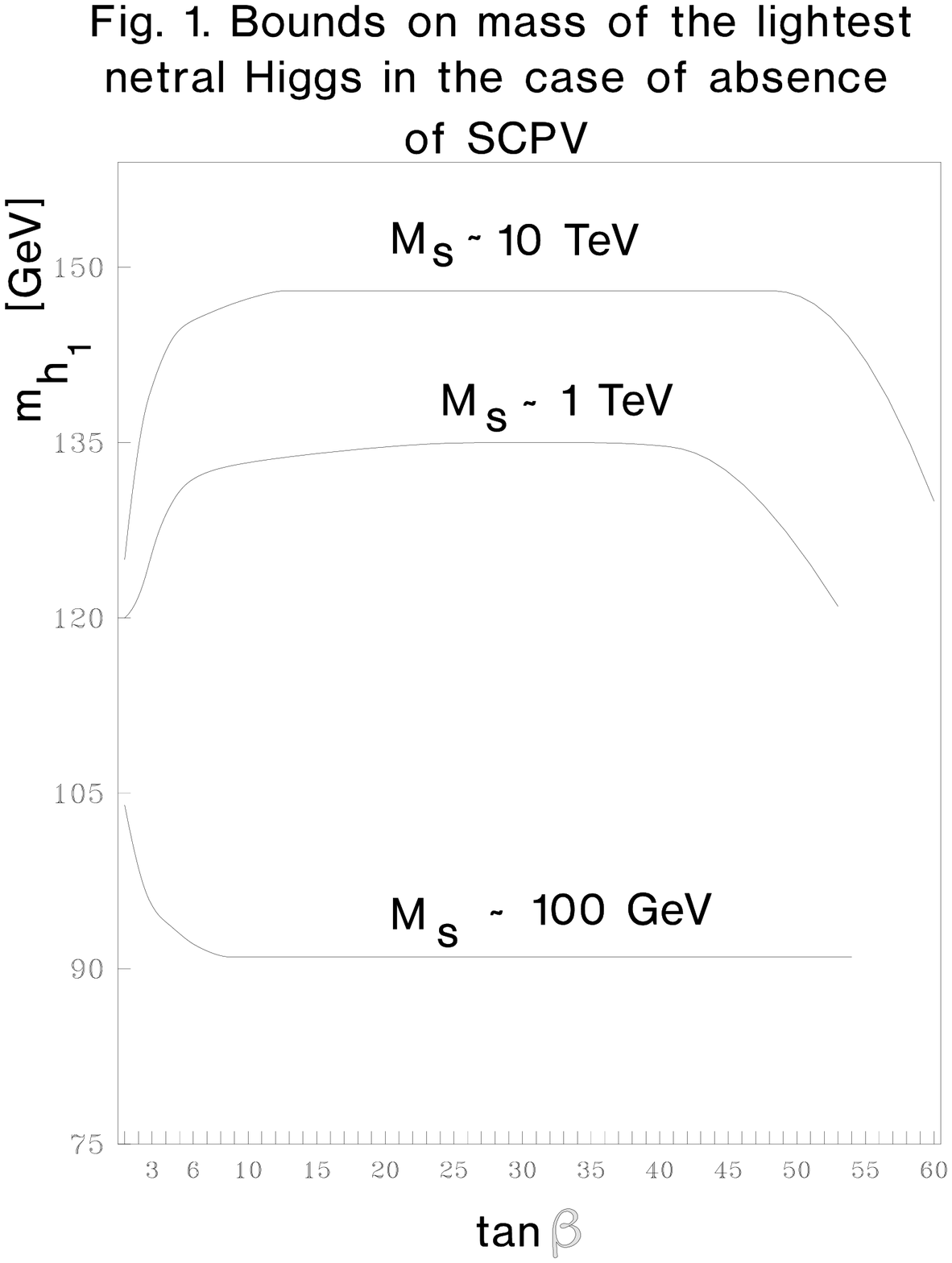}
\end{titlepage}
\begin{titlepage}
\setcounter{page}{0}
\epsfbox{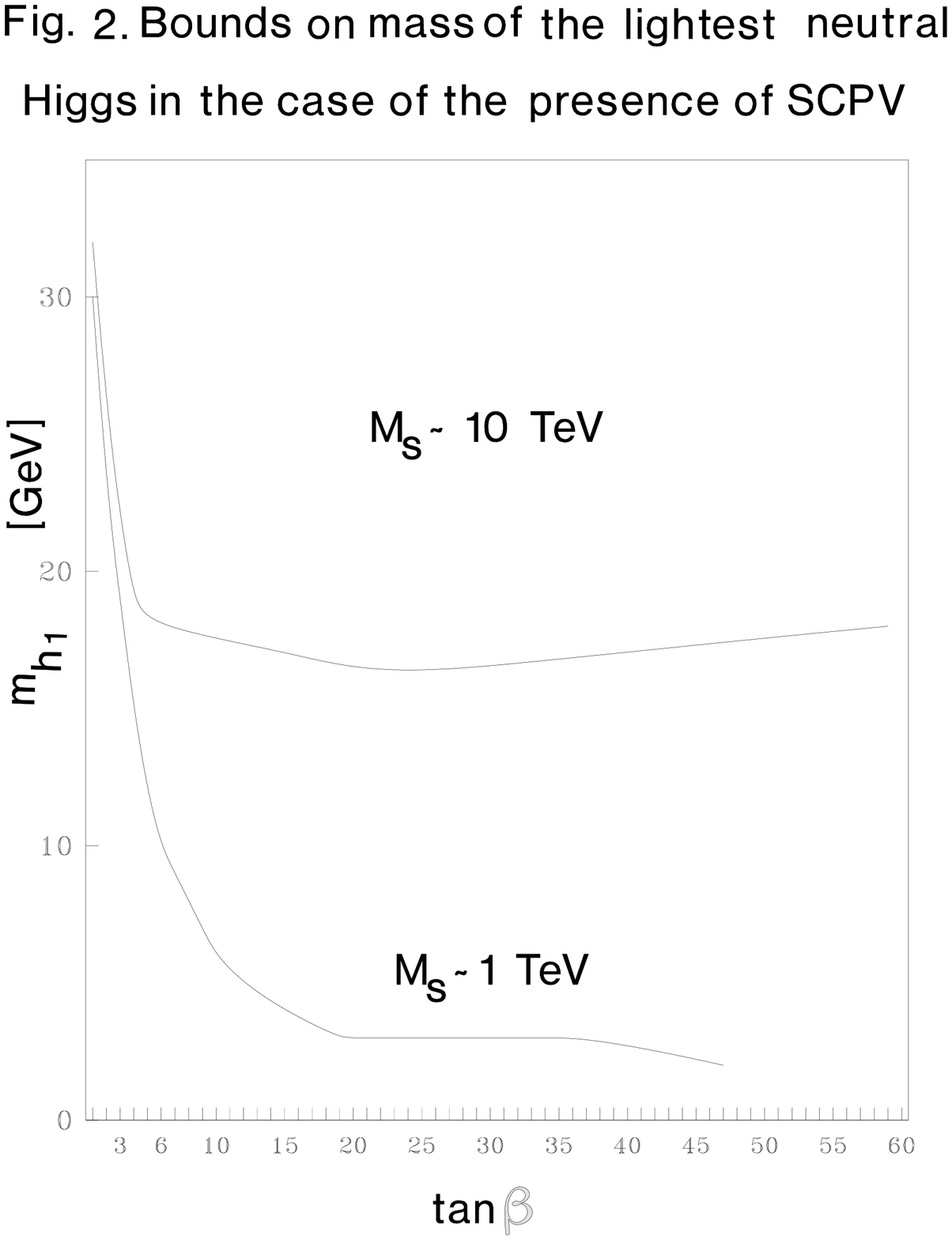}
\end{titlepage}
\begin{titlepage}
\setcounter{page}{0}
\epsfbox{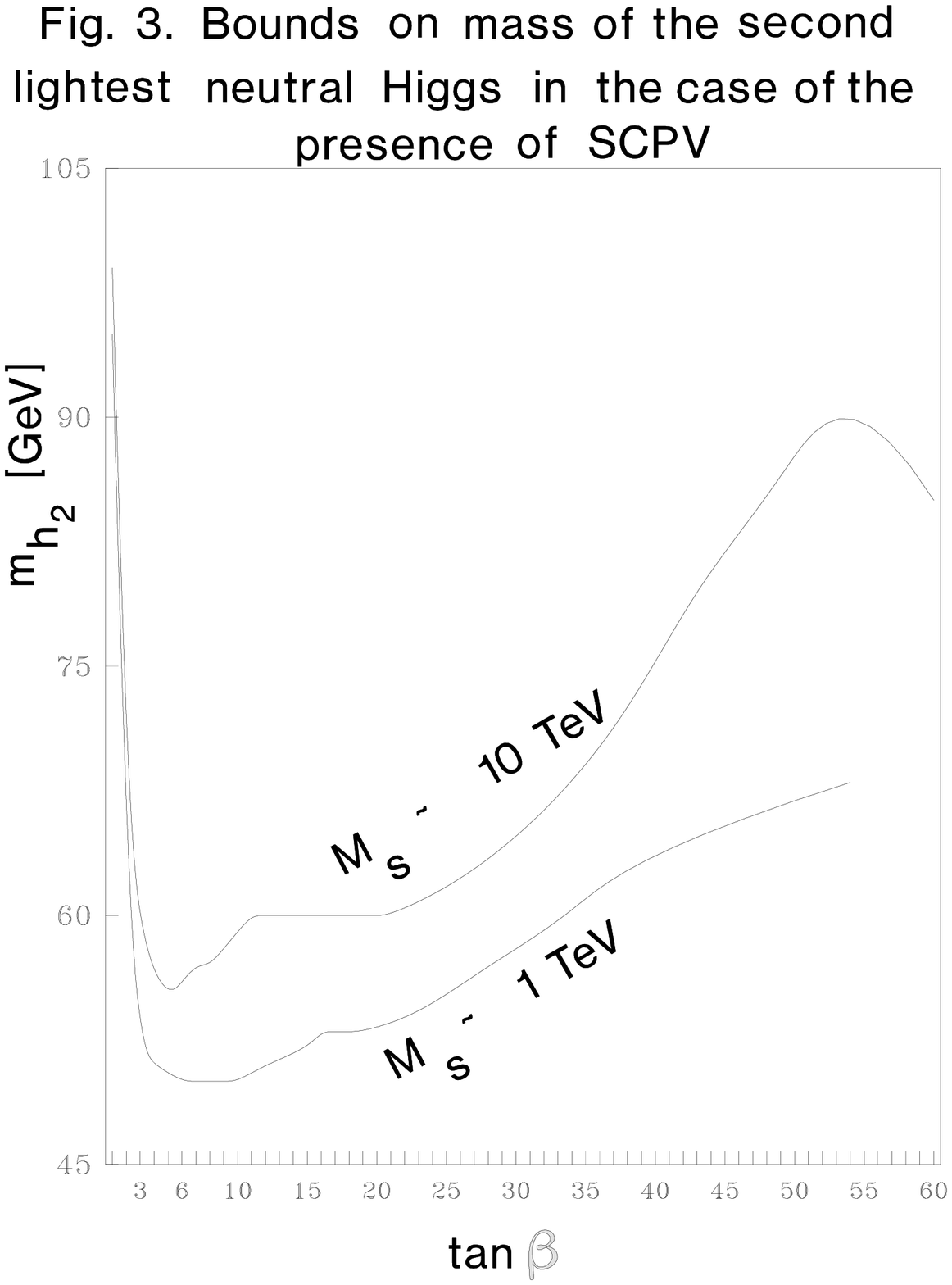}
\end{titlepage}
\begin{titlepage}
\setcounter{page}{0}
\epsfbox{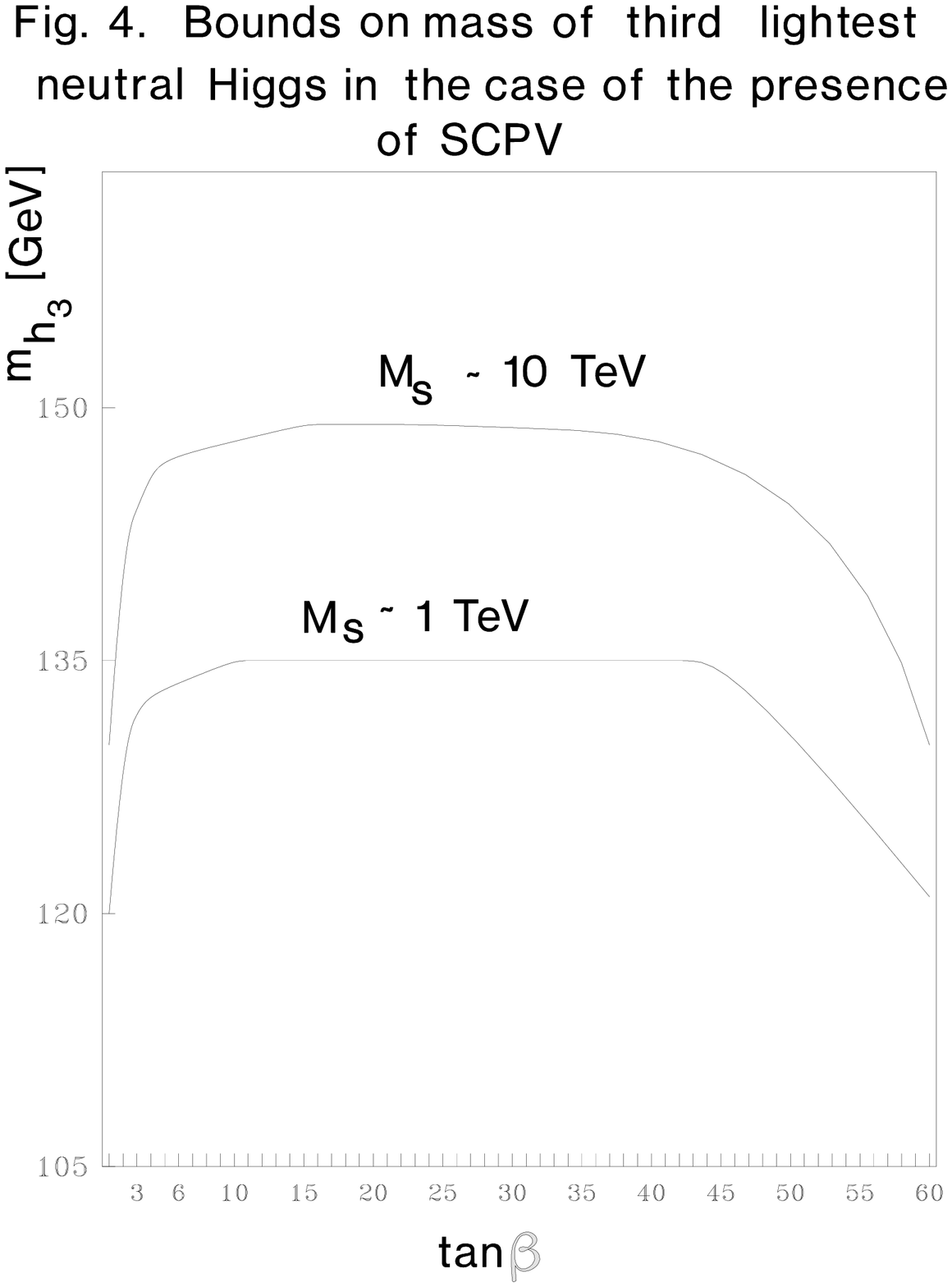}
\end{titlepage}
\begin{titlepage}
\setcounter{page}{0}
\epsfbox{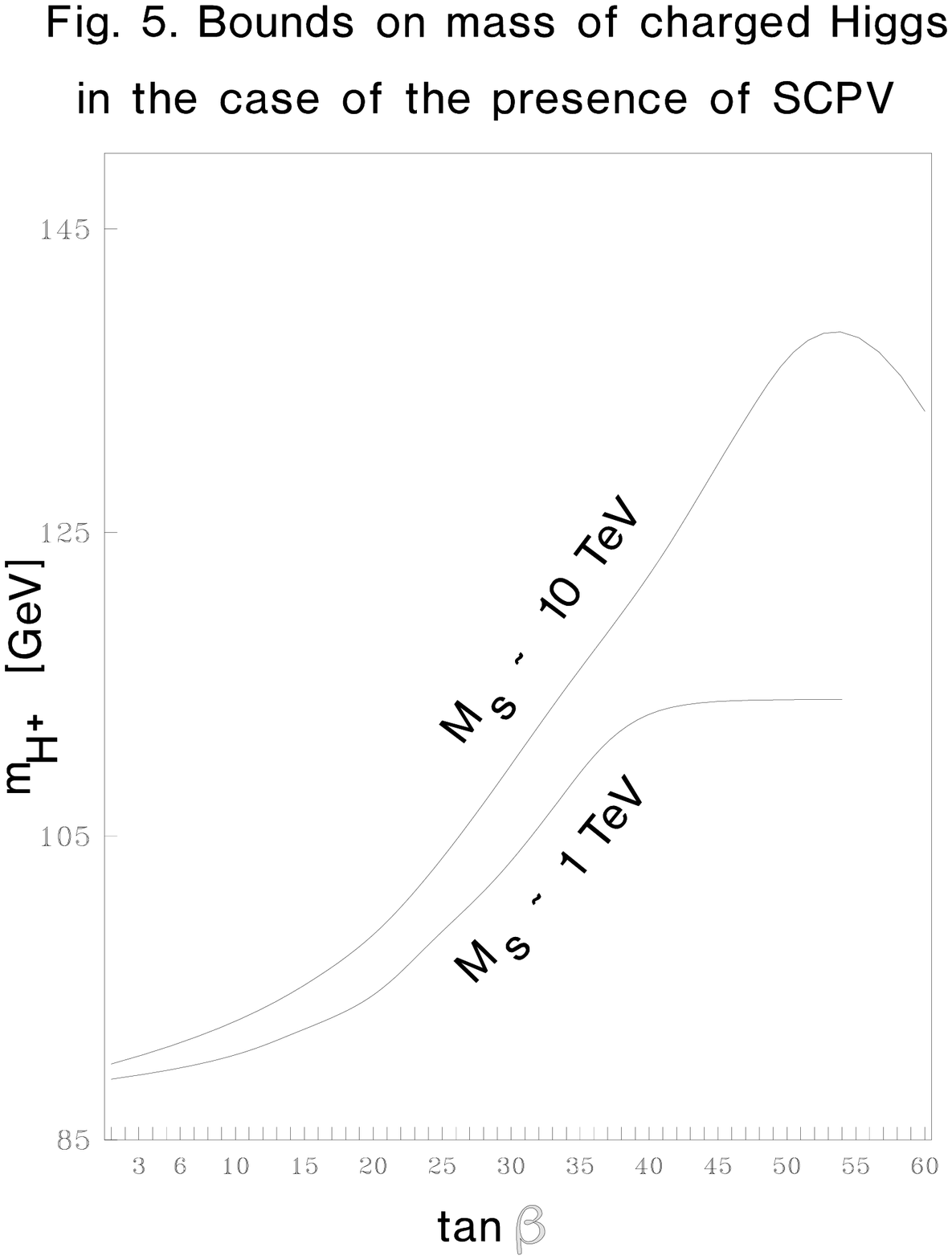}
\end{titlepage}

\newpage
\begin{thebibliography}{99}
\bibitem{1} H. P. Nilles, M. Srednicki, D. Willer Phys. Lett. 120B, 366
(1983).
\bibitem{2} A. Pomarol, Phys. Lett B287, 331 (1992).
\bibitem{3} A. Pomarol, Phys. Rev., 47D, 273 (1993).
\bibitem{4} S. Bertolini, F. Visssani, Phys. Lett. B324, 164 (1994).
\bibitem{5} A. Soni, R. M. Xu, Phys. Rev. Lett. 69, 33 (1992).
\bibitem{6} T. Elliot, S. King, P. White, Phys. Lett. 305B, 71 (1993).
\bibitem{7} Y. Okado, Preprint KEK-TH-382 (1993).
\bibitem{8} U. Ellwanger et al. Preprint LPTHE Orsay 95-04 (HEP-PH 9502206).
\bibitem{9} J. Ellis et al. Phys. Rev. D39, 844 (1989).
\bibitem{10} L. Maiani et al. Nucl. Phys. B136, 115 (1978).
\bibitem{11} N. Cabibbo et al. Nucl. Phys. B158, 295 (1979).
\bibitem{12} B. Pendlton, G. Ross, Phys. Lett. 98B, 221 (1981).
\bibitem{13} S. Dimopoulos, Proceedings of the XXVII Intern. Conf. on HEP
(Glasgow, 1994), p. 93.
\bibitem{14} H. Asatrian, A. Ioannisyan, S. Matiyan, Z. Phys. C61, 265 (1994).
\bibitem{15} H. P. Nilles, Preprint MPI-Ph/93-84 (1993).
\bibitem{16} N. Nilles, Phys. Rep. 110, 1 (1984).
\bibitem{17} R. Barbiery, S. Ferrara, C. Savoy, Phys. Lett. 119B, 343 (1982).
\bibitem{18} P. Nath, R. Arnowitt, A. Chamseddine, Phys. Rev. Lett. 49B,
970 (1982)
\bibitem{19}  L. Ibanez, Phys. Lett. 118B, 73 (1982).
\bibitem{20} L. Ibanez, Nucl. Phys. 218B, 514 (1983).
\bibitem{21} S. Soni, H. Weldon, Phys. Lett., 216B, 215 (1983).
\bibitem{22} P.H.Chankowski,S.Sikorski,J.Rosiek Phys. Lett 281B 100 (1992)
\bibitem{23} S. P. Martin, M. T. Vaughn, Preprint NUB-3081 93T
(HEP-PH 9311340).
\bibitem {24} H.B.Jensen Proceedings of the XXVII Intern. Conf. on HEP
(Glasgow, 1994), p. 3.
\bibitem{25} M. Bando et al. Mod. Phys. Lett. A7, 3379 (1992).
\bibitem{26} M. Pohl, Proceedings of the XXVII Intern. Conf. on HEP
(Glasgow, 1994), p. 107.
\end{thebibliography}
\end{document}